\documentclass{article}

\usepackage[right= 0.75in, left= 0.75in, top=1.05in, bottom=1.15in]{geometry}

\usepackage{lineno,hyperref}
\modulolinenumbers[5]

\usepackage{color}

\usepackage{amsmath}
\usepackage{amsfonts}
\usepackage{bm} 
\usepackage{booktabs} 
\usepackage{array} 
\usepackage{paralist} 
\usepackage{verbatim} 
\usepackage{subfig} 
\usepackage[super,sort&compress,comma]{natbib} 
\usepackage[utf8]{inputenc}
\usepackage[affil-it,auth-sc]{authblk}
\usepackage{graphicx}
\usepackage{amsmath}

\DeclareMathOperator{\divg}{\mbox{div}} 
\newcommand{\mb}[1]{\mathbf{#1}} 

\newcommand{\mc}[1]{\mathcal{#1}} 
\newcommand{\wt}[1]{\widetilde{#1}} 

\author[1]{Alessandro Lucantonio}
\author[1]{Antonio DeSimone\thanks{Corresponding author -- e-mail address:\,\texttt{desimone@sissa.it}.}}
\affil[1]{\small{SISSA--International School for Advanced Studies, via Bonomea 265, 34136 Trieste, Italy.}}

\title{Coupled swelling and nematic reordering in liquid crystal gels}

\date{}

\begin{document}
	
\maketitle
	
\begin{abstract}
We derive a multiphysics model that accounts for network elasticity with spontaneous strains, swelling and nematic interactions in liquid crystal gels (LCGs). 
We discuss the coupling among the various physical mechanisms, with particular reference to the effects of nematic interactions on chemical equilibrium and that of swelling on the nematic-isotropic transition. Building upon this discussion and using numerical simulations, we explore the transient phenomena involving concurrent swelling and phase transition in LCGs subject to a temperature change. Specifically, we demonstrate separation in time scales between solvent uptake and phase change, in agreement with experiments, which determines a kinetic decoupling between shape and volume changes. Finally, we discuss possible applications in the context of microswimmers, where such a kinetic decoupling is exploited to achieve non-reciprocal actuation and net motion in Stokes flow.
\end{abstract}

\section{Introduction}

Liquid crystal gels (LCGs) are fluid-swollen elastomers that contain rod-like molecules called \textit{mesogens}. These hybrid materials combine rubber elasticity of the cross-linked polymer network with the electro-optical properties of liquid crystals associated with the interacting mesogens. Hence, for instance, mesogens can be re-oriented through electric fields\cite{Urayama2006} or mechanical strain\cite{Kundler1995}, thus conferring a variable anisotropy to the solid, although flow is hampered because of the cross-linking.  Like their non-mesogenic counterpart, \textit{i.e.}~polymer gels, LCGs  can swell or shrink (anisotropically) in response to changes in the chemical potential of the swelling fluid. Additionally, variations in the temperature of the environment alter the nematic order, which in turn affects both the anisotropy and the magnitude of volume changes\cite{Urayama2005a}. Reciprocally, swelling modifies the strength of nematic interactions and thus influences nematic order. 

These coupled effects may be exploited in applications, where the greater compliance of LCGs with respect to dry liquid crystal elastomers (LCEs) is advantageous, since it allows for an easier manipulation of the director orientation. Specifically, LCEs have been proposed as artificial muscles\cite{Hebert1997}, light-activated microscopic walkers\cite{Zeng2015,DeSimone2015,Gelebart2017} and swimmers\cite{Huang2015}, substrates for stretchable electronic devices\cite{Ware2016}, and for shape morphing of thin sheets\cite{Aharoni2014,Mostajeran2015}, where the orientation of the director may be appropriately patterned to produce a predefined three-dimensional configuration, upon thermal activation of the nematic-isotropic transition. 
 
Previous theoretical and experimental reports have mainly focused on LCEs, while those regarding LCGs have essentially studied equilibrium problems\cite{Urayama2003,Urayama2005,Urayama2005a,Urayama2007a,Cheewaruangroj2015,Zakharov2017}. In particular, Cheewaruangroj and Terentjev\cite{Cheewaruangroj2015} have modeled the equilibrium phase separation in stretched nematic gels by combining anisotropic Gaussian elasticity\cite{Bladon1994,DeSimone2009} with Flory-Huggins mixing energy\cite{FloryBook1953}, without accounting for the energy of nematic interactions. In such a model the nematic-isotropic transition is accounted for via a prescribed dependence of the order degree on temperature, which may be recovered by fitting experimental data. The study of equilibrium phase separation in nematic gels was  recently resumed in the context of nemato-elastic shells by Zakharov and Pismen\cite{Zakharov2017}, who extended the previous model with a nematic energy.

Here, we are interested in investigating the transient coupled phenomena that involve anisotropic swelling, nematic-isotropic phase transition and nematic reordering in LCGs. First, we formulate a non-linear, coupled model for LCGs based on anisotropic Gaussian elasticity, Flory-Huggins solvent-polymer mixing theory, and Landau-de Gennes theory of nematic interactions\cite{WarnerBook}. As a fundamental coupling mechanism in this model, we introduce a dependence on solvent concentration of the strength of the interactions among the mesogens. We then discuss how this coupling affects the chemical equilibrium and the nematic-isotropic transition. To complete the theoretical setting, we present a variational formulation of the model, which is then used for the finite element implementation of the weak-form governing equations.

Inspired by experiments on the swelling dynamics of a LCG\cite{Urayama2006a}, we perform numerical simulations involving a freely swelling nematic gel subject to a temperature increase, to gather insight on the coupling between swelling and nematic order. We analyze the transient evolution of deformation, chemical potential, and nematic order, to rationalize experimental observations. In particular, we observe that  nematic-isotropic phase transition and swelling occur on separate time scales, in agreement with experiments. The existence of separate time scales in the evolution of nematic order and solvent uptake, which determines a non-reciprocal time evolution of shape under periodic actuation, naturally suggests to explore possible applications of LCGs in the context of swimming at microscopic scales. We explore the feasibility of this concept through numerical experiments on a prototype of nematic gel swimmer. 

\section{A coupled model for LCGs}

In this section, we derive a coupled model accounting for  elasticity, solvent migration and nematic interactions in swollen LCGs.  
We also present a weak formulation of the governing equations, which allows for a finite element implementation of the model. 

\subsection{Kinematics}

We consider a LCG consisting of a \textit{monodomain} LCE swollen with an \textit{isotropic} solvent.
Let $\mc{B}_{\rm d}$ be the dry configuration of the LCG in its isotropic phase, $\mc{B}$ the reference configuration and $\mc{B}_t$ the current configuration of the system at time $t$. The dry configuration, consisting of the solvent-free, undeformed LCE, serves as a reference to measure stretching of the polymer chains, swelling and the associated polymer-solvent mixing energy. The reference configuration models a \textit{stress-free} (swollen) preparation state of the LCG, which generally contains a certain amount of solvent and where the elastomer may be in  its either isotropic or nematic phase. In practice, it is convenient to measure displacements and strains with respect to the reference configuration, which constitutes the computational domain in numerical simulations and coincides with the initial configuration of the system. 

The \textit{state} of the LCG at each time is identified by the \textit{deformation} from $\mc{B}$ with deformation gradient $\mb{F}$, the \textit{solvent concentration} $c$ per unit reference volume, the current \textit{director orientation} $\mb{N}=\mb{n}\otimes\mb{n}$ and the \textit{degree of nematic order} $Q$. As usual, the deformation gradient $\mb{F}$ is related to the \textit{displacement field} $\mb{u}$ by $\mb{F}=\mb{I}+\nabla\mb{u}$, where $\mb{I}$ is the identity tensor. In the reference configuration $\mc{B}$ the (initial) director orientation is $\mb{N}_{\rm o}=\mb{n}_{\rm o}\otimes\mb{n}_{\rm o}$, the (initial) solvent concentration per unit reference volume is $c_{\rm o}$ and the (initial) scalar order parameter is $Q_{\rm o}$. The reference configuration is ideally reached from $\mc{B}_{\rm d}$ through an \textit{anisotropic free swelling} (see also Section~\ref{sec:freeswell}) characterized by the deformation gradient $\mb{F}_{\rm o}$. Consistent with this kinematical setting, we decompose the deformation gradient $\mb{F}_{\rm d}$ from the dry state as
\begin{align}
	\mb{F}_{\rm d} = \mb{F}\mb{F}_{\rm o}\,, \quad \mb{F}_{\rm o} = \mb{F}_{\rm os}\mb{F}_{\rm on}\,, \label{eq:1}
\end{align}
where $\mb{F}_{\rm os}=J_{\rm o}^{1/3}\mb{I}$ is an isotropic free swelling with volume ratio equal to $J_{\rm o}$ and $\mb{F}_{\rm on}$ is \textit{spontaneous nematic stretch} associated with $(Q_{\rm o},\mb{N}_{\rm o})$ such that\cite{Biggins2012}
\begin{align}
	\mb{L}_{\rm o}=\mb{F}_{\rm on}\mb{F}_{\rm on}^{\rm T}=\mb{L}(Q_{\rm o},\mb{N}_{\rm o})  = r_{\rm o}^{2/3}\mb{N}_{\rm o}+r_{\rm o}^{-1/3}(\mb{I}-\mb{N}_{\rm o})\,,  \label{eq:3}
\end{align}
with $r_{\rm o} = r(Q_{\rm o})$. Here, $r(Q)$ is the anisotropy of the polymer network backbone for a given $Q$ 
\begin{align}
r(Q) = \frac{1+2aQ}{1-aQ}\,, \label{eq:r}
\end{align}
where the parameter $0<a\leq 1$ accounts for mismatches between the degree of order of mesogens and that of the backbone in side-chain nematic elastomers\cite{WarnerBook}. 

Following DeSimone and Teresi\cite{DeSimone2009}, we introduce the  \textit{elastic} stretch tensor from the dry state as
\begin{align}
\mb{B}_{\rm de} = \mb{F}_{\rm d}\mb{F}_{\rm d}^{\rm T}\mb{L}^{-1} = \mb{F}\mb{B}_{\rm o}\mb{F}^{\rm T}\mb{L}^{-1}\,, \quad \mb{B}_{\rm o} = \mb{F}_{\rm o}\mb{F}_{\rm o}^{\rm T}=J_{\rm o}^{2/3}\mb{L}_{\rm o}\,. \label{eq:Bde}
\end{align}
where $\mb{L}=\mb{L}(Q,\mb{N})$. As we will see in the next Section, this stretch tensor identifies the part of $\mb{F}_{\rm d}\mb{F}_{\rm d}^{\rm T}$ that is effective in storing elastic energy.
In the following, we assume that the director field is simply convected by the motion of the polymer network, as it is the case for nematic glasses, \textit{i.e.} $\mb{n}=\mb{F}\mb{n}_{\rm o}/|\mb{F}\mb{n}_{\rm o}|$. 

As usual in the theory of polymer gels, both the polymer matrix and the solvent are assumed to be separately incompressible, so that their volumes are additive. Hence, we can express the local volume change $J=\det\mb{F}$ of the gel in terms of the change $\Omega(c-c_0)$ in solvent volume fraction as\cite{Lucantonio2013}
\begin{align}
J = 1+\Omega (c-c_{\rm o})\,, \label{eq:swellconstr}
\end{align}
where $\Omega$ is the solvent molar volume. Let us notice that the solvent volume fraction with respect to the dry state is $\phi_{\rm d} = 1/(J_{\rm o}J)$.

\subsection{Free energy and constitutive equations}

We assume that the Helmholtz free energy $\Psi$ per unit reference volume of the LCG consists of three contributions: the elastic energy $\Psi_{\rm e}$ of the polymer network, the solvent-polymer mixing energy $\Psi_{\rm m}$, and the nematic energy $\Psi_{\rm n}$ of liquid crystals. The total  free energy of the system is then given by
\begin{align}
\begin{split}
\mc{E}[\mb{F},c,Q,p] &= \int_{\mc B}{(\Psi(\mb{F},c,Q)-p g(\mb{F},c))}\,, \label{eq:totenergy}
\end{split}
\end{align} 
where $\Psi(\mb{F},c,Q) = \Psi_{\rm e}(\mb{F},Q) + \Psi_{\rm m}(c)
+ \Psi_{\rm n}(c,Q)$ and where the last term enforces the constraint $g(\mb{F},c)= J - 1-\Omega (c-c_{\rm o})=0$ coming from eq.~\eqref{eq:swellconstr} \textit{via} the Lagrange multiplier $p$. Here, we have assumed specific functional dependences of the various energy terms on the state variables, which are justified by the choices for the representation forms of these energies we are going to perform next.

Following previous works\cite{DeSimone2009,Cheewaruangroj2015} on elastic energies for LCEs, we take the elastic free energy density per unit reference volume as\footnote{Here and in what follows ``$\cdot$'' denotes the scalar product between both tensors and vectors.}
\begin{align}
\Psi_{\rm e}(\mb{F},Q) = \frac{1}{J_{\rm o}}\frac{G}{2}\left[\mb{I}\cdot\mb{B}_{\rm de} - 3 - \log (\det \mb{B}_{\rm de})\right]\,, \label{eq:elastic} 
\end{align}
which is minimized by every $\mb{F}=\mb{L}^{1/2}\mb{Q}\mb{L}_{\rm o}^{-1/2}/J_{\rm o}^{1/3}$, with $\mb{Q}$ an arbitrary rotation. 
Here, $G$ is the shear modulus of the dry polymer network that, as is known from rubber elasticity\cite{FloryBook1953},  depends linearly on the temperature $T$: $G=Nk_{\rm B}T$, where $N$ is the  density of polymer chains per unit dry volume and $k_{\rm B}$ is the Boltzmann constant. We are regarding the energy density as a function of $\mb{F}$ and $Q$ through the relations \eqref{eq:3}-\eqref{eq:r} and \eqref{eq:Bde}. Equation \eqref{eq:elastic} comes from a rewriting\cite{DeSimone2009} of the well-known ``Trace formula''\cite{WarnerBook}, where the trace term is $\mb{I}\cdot\mb{B}_{\rm o}\mb{F}^{\rm T}\mb{L}^{-1}\mb{F} = \mb{I}\cdot\mb{B}_{\rm de}$. Here, the reference nematic stretch $\mb{L}_{\rm o}$  contained in the original Trace formula has been replaced by $\mb{B}_{\rm o}$, because in addition to $\mb{L}_{\rm o}$ the polymer chains also suffer an isotropic stretch due to the initial swelling in the transformation from the dry isotropic configuration to the reference one. The logarithm term is typical of compressible Neo-Hookean elasticity.
The energy density \eqref{eq:elastic} could be easily extended with anisotropic corrections\cite{DeSimone2009} to account for the preferential nematic orientation imprinted during the cross-linking process. While quantitatively affecting  equilibrium, these corrections do not qualitatively alter  the coupling structure between swelling and nematic order, which is the scope of the present analysis, and are thus hereafter neglected.

To quantify the entropy and enthalpy related to mixing of solvent and polymer network, we employ the Flory-Huggins theory\cite{FloryBook1953}, which prescribes the following representation for the mixing free energy per unit reference volume:
\begin{align}
\Psi_{\rm m}(c)=\frac{1}{J_{\rm o}}\frac{R_{\rm g}T}{\Omega}\left(\Omega J_{\rm o} c\log\frac{\Omega J_{\rm o} c}{1+\Omega J_{\rm o} c}+\chi\frac{\Omega J_{\rm o} c}{1+\Omega J_{\rm o} c}\right)\,, \label{eq:flory}
\end{align}
where $\chi$ is a temperature-dependent dimensionless measure of the enthalpy of mixing that is determined by the chemical affinity between solvent and polymer and $R_{\rm g}$ is the universal gas constant. Mixing interactions between polymer and mesogens and between solvent and mesogens could also be accounted for in a similar fashion. Both the elastic and the mixing energy are naturally introduced as densities with respect to the dry configuration; the $1/J_{\rm o}$ factor in eqs.~\eqref{eq:elastic}-\eqref{eq:flory} accounts for the change of density when they are pushed-forward to the reference configuration.

For the nematic energy describing the interaction among the mesogens, we take the fourth-order Landau-de Gennes expansion\cite{WarnerBook} 
\begin{align}
&\Psi_{\rm n}(c,Q)  \!= \!\frac{\xi}{J_{\rm o}}\left(\frac{1}{2} A(c,T) Q^2 \!-\!\frac{1}{3} B Q^3 \!+\frac{1}{4} C Q^4\right), \label{eq:nem}
\end{align}
where $\xi$ is the volume fraction of mesogens in the dry state and $A$, $B$, and $C$ are non-negative. Near the nematic-isotropic transition, where this expansion is valid in principle, $B$ and $C$ have a weaker dependence on temperature than $A$ and are thus assumed to be $T$-independent. Terms proportional to the gradient of the order parameter may be added and could be especially relevant in the presence of defects of the director orientation. In the representation \eqref{eq:nem} of the nematic free energy the coefficient $A$ models the strength of the local nematic interactions, which we also expect to depend on swelling in addition to the ordinary dependence on temperature in liquid crystals.  Indeed, dilution of the mesogens causes the average distance among them to increase, thus reducing the magnitude of distance-dependent pairwise interactions (such as van der Waals forces). According to this physical picture, we prescribe for $A$ the following formula
\begin{align}
A(c,T)=A_{\rm o}(T-T_{\rm s}(c))\,, \quad T_{\rm s}(c)=T_{\rm sd}\xi f(\phi_{\rm d})\,, \quad \phi_{\rm d} = \frac{1}{1+\Omega J_{\rm o}c}\label{eq:A}
\end{align}
where we have modeled the aforementioned dilution effect as a solvent concentration-dependent shift\cite{Cheewaruangroj2015,Zakharov2017,Zakharov2017a} in the characteristic temperature $T_{\rm s}$, which is related to the transition temperature, as will be discussed in Section~\ref{sec:freeswell}, in the context of homogeneous equilibrium states. Here, $A_{\rm o}$ is a characteristic value for the strength of the nematic interactions, $T_{\rm sd}\xi$ is the value of $T_{\rm s}$ for the dry LCE and $f$ is an \textit{increasing} function of $\phi_{\rm d}$ such that $f(1)=1$. According to eq.~\eqref{eq:A}, at a given temperature $T<T_{\rm sd}\xi$, swelling ($\phi_{\rm d}$ \textit{decreases}) makes $T_{\rm s}$ decrease; hence, $A$ decreases in modulus, so that weaker interactions among the mesogens tend to stabilize the isotropic phase.

In addition to the effect of swelling on the strength of nematic interaction, the presence of cross-links in LCEs and LCGs influences nematic order.
Indeed, the dependence of the elastic energy $\Psi_{\rm e}$ on $Q$ also plays a role, in the form of the elastic component of the Cauchy stress\cite{Noselli2016} $\mb{T}_{\rm e}=(\partial \Psi_{\rm e}/\partial \mb{F})\mb{F}^{\rm T}/J$, in determining the equilibrium degree of order through the term
\begin{align}
	&\frac{\partial \Psi_{\rm e}}{\partial Q}= \frac{1}{2}\frac{a r^{-2/3}}{(1-aQ)^2}\mb{L}\mb{T}_{\rm e}\cdot[\mb{I}-(1+2r^{-1})\mb{N}]\,,\label{eq:partialQ1}
\end{align}
which contributes to the stationarity condition of the \textit{total} free energy $\Psi$ with respect to $Q$.
However, notice that a spherical elastic stress causes the derivative in eq.~\eqref{eq:partialQ1} to vanish because $\mb{L}\cdot[\mb{I}-(1+2r^{-1})\mb{N}]=0$, so that elasticity does not directly affect the equilibrium degree of order. This is the case, for instance, of stress-free states that we analyze in Section~\ref{sec:freeswell}, such as the reference configuration we have chosen. For a liquid crystal, the equilibrium degree of order is simply characterized by the vanishing of the derivative
\begin{align}
	&\frac{\partial \Psi_{\rm n}}{\partial Q} = \frac{\xi}{J_{\rm o}}Q\left(A(c,T)-BQ+CQ^2\right) \label{eq:partialQ2}\,.
\end{align}
The isotropic phase $Q=0$ is always a stationary point of $\Psi_{\rm n}$; at $T=T_{\rm u} = T_{\rm s} + B^2/(4AC)$ another equilibrium point appears for $\Psi_{\rm n}$, which persists for $T<T_{\rm u}$. 

Now, having provided the representations for the energy contributions \eqref{eq:elastic},\eqref{eq:flory}, and \eqref{eq:nem} that form the free energy density $\Psi$ defined after eq.~\eqref{eq:totenergy}, we can derive the constitutive equations for the (Piola-Kirchhoff) stress $\mb{S}$ and the chemical potential $\mu$ of the solvent within the gel. Precisely, these equations are 
\begin{align}
&\mb{S} = \frac{\partial \Psi}{\partial \mb{F}} +p\frac{\partial g}{\partial \mb{F}} = \frac{G}{J_{\rm o}}(\mb{L}^{-1}\mb{F}\mb{B}_{\rm o}-\mb{F}^{-\rm T})-p\mb{F}^\star\,, \label{eq:stress}  \\
&\mu = \frac{\partial \Psi}{\partial c} +p\frac{\partial g}{\partial c} = \Omega(p-\Pi-p_{\rm n})\,, \label{eq:chempot}
\end{align}
where $\mb{F}^\star = J\mb{F}^{-\rm T}$ and we have introduced the \textit{osmotic pressure} $\Pi$ of the solvent-polymer mixture and the \textit{nematic pressure} $p_n$ defined as
\begin{align}
&\Pi = -\frac{R_{\rm g}T}{\Omega}\left(\log\frac{\Omega J_{\rm o}c}{1+\Omega J_{\rm o}c}+\frac{1}{1+\Omega J_{\rm o}c}+\chi\frac{1}{(1+\Omega J_{\rm o}c)^2}\right)\,, \\
&p_n = -\frac{1}{2}\frac{\xi^2 A_{\rm o}T_{\rm sd}}{(1+\Omega J_{\rm o}c)^2}f'\left(\frac{1}{1+\Omega J_{\rm o}c}\right)Q^2\,, \label{eq:nempress}
\end{align}
respectively. The constitutive equation \eqref{eq:chempot} can be physically interpreted as the definition of $p$ as the \textit{total pressure} within the LCG. Indeed, given that the solvent pressure is $p_{\rm s} = \mu/\Omega$, being an incompressible fluid, we can recast eq.~\eqref{eq:chempot} as
\begin{align}
p = \Pi + p_{\rm s} + p_{\rm n}.
\end{align}
Furthermore, as it is evident from eq.~\eqref{eq:chempot}, temperature variations induce changes in the chemical potential both \textit{directly}, through the dependence on $T$ of the mixing parameter $\chi$, and \textit{indirectly},  through the temperature-dependent scalar order parameter $Q$. This is a fundamental coupling mechanism between swelling and nematic order that we will see in action in the numerical examples presented in Section~\ref{sec:numerics}. On the other hand, deviations from chemical equilibrium caused by changes in the chemical potential of the environment  drive swelling, which in turn varies $T_{\rm s}$ and could thus activate the nematic-isotropic transition at a fixed temperature. This effect has been recently demonstrated\cite{Boothby2017} and exploited in swelling-triggered (chemoresponsive) isothermal LCG actuators.

\subsection{Anisotropic free swelling and nematic-isotropic transition; characterization of the reference configuration}
\label{sec:freeswell}

We consider a class of equilibrium configurations the system may attain by swelling anisotropically in the absence of body forces or constraints. In the present context, these \textit{anisotropic free swelling} configurations are \textit{homogeneous} states characterized by zero stress ($\mb{S}=\mb{0}$), chemical equilibrium ($\mu=\mu_{\rm e}$) with an external solvent at chemical potential $\mu_{\rm e}$, and by the following representation of the deformation gradient
\begin{align}
\mb{F} = \mb{R}(\lambda_{\parallel}\mb{N}_{\rm o}+\lambda_{\perp}(\mb{I}-\mb{N}_{\rm o})) \label{eq:anisoswellF}
\end{align}
where $\mb{R}$ is a rotation. The (Lagrangian) principal stretches are aligned along the directions parallel and orthogonal to the reference director $\mb{N}_{\rm o}$. For instance, if $\mb{N}_{\rm o}$ lies in the plane $\mb{e}_1$-$\mb{e}_2$ and is such that $\mb{F}\mb{e}_1=\lambda\mb{e}_1$, eq.~\eqref{eq:anisoswellF} represents a (generalized simple) shear in such plane with a superimposed extension. In this case, the stretch along $\mb{e}_1$ is given by
\begin{align}
\lambda = \sqrt{[\lambda_{\parallel}(\mb{N}_{\rm o})^2_{11}+\lambda_{\perp}(1-(\mb{N}_{\rm o})^2_{11})]^2+[(\lambda_{\parallel}-\lambda_{\perp})(\mb{N}_{\rm o})_{12}]^2}\,,
\end{align}
while the rotation occurs about the $\mb{e}_3$ axis with an angle $\theta$ such that
\begin{align}
\tan\theta = \frac{(\lambda_{\parallel}-\lambda_{\perp})(\mb{N}_{\rm o})_{12}}{\lambda_{\parallel}(\mb{N}_{\rm o})^2_{11}+\lambda_{\perp}(1-(\mb{N}_{\rm o})^2_{11})}\,, 
\end{align}
with $(\mb{N}_{\rm o})_{ij} = \mb{N}_{\rm o}\cdot\mb{e}_{i}\otimes\mb{e}_j$.

Taking into account the representation \eqref{eq:anisoswellF}, the conditions of zero stress, chemical equilibrium and the optimal degree of order specialize as
\begin{align}
&\frac{G}{J_{\rm o}}\left(J_{\rm o}^{2/3}r^{-2/3}\lambda_{\parallel}r_{\rm o}^{2/3}-\frac{1}{\lambda_{\parallel}}\right)-p\lambda_{\perp}^2 = 0\,, \label{eq:stressfree1} \\
&\frac{G}{J_{\rm o}}\left(J_{\rm o}^{2/3}r^{1/3}\lambda_{\perp}r_{\rm o}^{-1/3}-\frac{1}{\lambda_{\perp}}\right)-p\lambda_{\parallel}\lambda_{\perp} = 0\,, \label{eq:stressfree2}  \\
&\Omega (p-\Pi-p_{\rm n})=\mu_{\rm e}\,, \\
&\Pi = -\frac{R_{\rm g}T}{\Omega}\left(\log\frac{J_{\rm o}\lambda_{\parallel}\lambda_{\perp}^2-1}{J_{\rm o}\lambda_{\parallel}\lambda_{\perp}^2}+\frac{1}{J_{\rm o}\lambda_{\parallel}\lambda_{\perp}^2}+\chi\frac{1}{(\lambda_{\parallel}\lambda_{\perp}^2J_{\rm o})^2}\right)\,, \label{eq:Pifree} 
\end{align}
\begin{align}
&p_{\rm n}=-\frac{1}{2}\frac{ \xi^2 A_{\rm o}T_{\rm sd}}{(J_{\rm o}\lambda_{\parallel}\lambda_{\perp}^2)^2}f'\left(\frac{1}{J_{\rm o}\lambda_{\parallel}\lambda_{\perp}^2}\right)Q^2\,, \label{eq:pnfree} \\
&Q=0 \quad \mbox{or} \quad Q=\frac{B\pm\sqrt{B^2-4A((J_{\rm o}\lambda_{\parallel}\lambda_{\perp}^2 -1)/\Omega J_{\rm o},T)C}}{2C}\,, \label{eq:freeswellQ}
\end{align}
where the constitutive equations \eqref{eq:stress}-\eqref{eq:nempress} and the  relation $c_{\rm o}=(J_{\rm o}-1)/\Omega J_{\rm o}$ have been used along with the partial derivatives with respect to $Q$ as given by eqs.~\eqref{eq:partialQ1}-\eqref{eq:partialQ2}.  In eq.~\eqref{eq:freeswellQ}, we have noticed from eqs.~\eqref{eq:partialQ1},\eqref{eq:stress} that $\partial \Psi_{\rm e}/\partial Q = 0$ when $\mb{S}=\mb{0}$. Elimination of $p$ between eqs.~\eqref{eq:stressfree1}-\eqref{eq:stressfree2} provides the equilibrium anisotropy ratio
\begin{align}
\alpha = \frac{\lambda_{\parallel}}{\lambda_{\perp}} = \sqrt{\frac{r}{r_{\rm o}}}\,,
\end{align}
which is a function of $Q$ only.

The set of equations \eqref{eq:stressfree1}-\eqref{eq:freeswellQ}  corresponds to the stationary points of $\Psi-\mu_{\rm e}c$. We now turn to the question of the nematic-isotropic transition temperature, which amounts to evaluating the stability of these points as a function of temperature. At the nematic-isotropic transition temperature two equal-energy minima exist. For liquid crystals this temperature is\cite{WarnerBook} $T_{\rm ni} = T_{\rm s} + 2B^2/(9A_{\rm o}C)$. In the special case $B=0$, a second-order phase transition occurs at $T=T_{\rm ni}=T_{\rm s}$. For nematic elastomers and gels, the transition temperature is generally affected by the presence of cross-links and mixing. 
Specifically, upon imposing that the total free energy density of the nematic phase is equal to that of the isotropic phase, we obtain the nematic free energy $\Psi_{\rm n}^{\rm nem}$ of the nematic phase as
\begin{align}
\Psi_{\rm n}^{\rm nem} = \Delta \Psi_{\rm m}+\Delta \Psi_{\rm e}\,,
\end{align}
where $\Delta \Psi_{\rm m}=\Psi_{\rm m}^{\rm iso}-\Psi_{\rm mix}^{\rm nem}$ and $\Delta \Psi_{\rm e}=\Psi_{\rm e}^{\rm iso}-\Psi_{\rm e}^{\rm nem}$. We notice that $\Delta \Psi_{\rm m} < 0$, because the decreased degree of order in the isotropic state with respect to the nematic one promotes swelling, which is associated to a reduction in the mixing free energy. We deduce that mixing \textit{favors} the transition, thus \textit{lowering} the value of the transition temperature for the gel with respect to $T_{\rm ni}$. On the contrary, elasticity may favor the nematic or the isotropic phase, according to the state at formation\cite{WarnerBook}.

Finally, we can characterize the reference configuration in terms of the initial swelling ratio $J_{\rm o}$ and the initial degree of order $Q=Q_{\rm o}$, by evaluating the above relations at $\lambda_{\parallel}=\lambda_{\perp}=1$. In particular, eqs.~\eqref{eq:stressfree1}-\eqref{eq:stressfree2} provide the initial pressure $p=p_{\rm o}$ as
\begin{align}
p_{\rm o} = \frac{G}{J_{\rm o}}(J_{\rm o}^{2/3}-1)\,.
\end{align}
Using this expression, the chemical equilibrium condition may be explicitly written as
\begin{align}
\begin{split}
&\log\frac{J_{\rm o}-1}{J_{\rm o}}+\frac{1}{J_{\rm o}}+\chi\frac{1}{J_{\rm o}^2}
+\frac{1}{2}\frac{ \Omega \xi^2 A_{\rm o}T_{\rm sd}}{J_{\rm o}^2 R_{\rm g}T_{\rm o}}f'\left(\frac{1}{J_{\rm o}}\right)Q_{\rm o}^2+ \\		
&\qquad +\frac{G\Omega}{J_{\rm o}R_{\rm g}T_{\rm o}}(J_{\rm o}^{2/3}-1)=\frac{\mu_{\rm o}}{R_{\rm g}T_{\rm o}}\,, \label{eq:equi1}
\end{split}
\end{align}
where $\mu_{\rm e}=\mu_{\rm o}$ is the initial chemical potential of the external solvent and $Q_{\rm o}$ is the stationary point of $\Psi_{\rm n}$ among the ones in eq.~\eqref{eq:freeswellQ} that corresponds to the stable phase at a given initial temperature $T_{\rm o}$.

\subsection{Weak formulation and variational setting}

To solve the governing equations of the coupled model using the finite element method, we start from their weak form. For the boundary conditions, we assume that the body is traction-free and in chemical equilibrium on the subset $\partial_{\mu}\mc{B}$ of the boundary with the external solvent whose chemical potential is $\mu_{\rm e}$ (on the rest of the boundary the solvent flux vanishes). We enforce this chemical equilibrium condition through the Lagrange multiplier $\eta$, which coincides with the normal solvent flux $\mb{h}\cdot\mb{m}$ on $\partial_{\mu}\mc{B}$, with $\mathbf{h}$ the solvent mass flux (measured in $\mbox{mol}/\mbox{m}^2\cdot\mbox{s}$) and $\mb{m}$  the outwards unit normal to $\partial \mc{B}$. For the free swelling of the cubic sample, symmetry conditions on $\mathbf{u}$ were imposed. Upon denoting by a superposed tilde the test fields, the weak formulation of the coupled model consists of the following equations:
\begin{itemize}
	\item balance of forces
		\begin{align}
			\int_{\mc B}{\mb{S}\cdot\nabla\wt{\mb{u}}} = 0\,, \label{eq:forcesweak}
		\end{align}
	\item balance of solvent mass
		\begin{align}
			\int_{\mc B}{(\dot{c}\tilde{\mu}-\mb{h}\cdot\nabla\tilde{\mu})}+\int_{\partial_\mu\mc B}{\eta\tilde{\mu}} = 0\,,
		\end{align}
	\item nematic equilibrium
		\begin{align}
			\int_{\mc B}{\frac{\partial \Psi}{\partial Q}\tilde Q} = 0\,, \label{eq:weaknem}
		\end{align}
	\item swelling constraint
		\begin{align}
			\int_{\mc B}{g(\mb{F},c)\tilde{p}}=0\,,
		\end{align}
	\item chemical equilibrium on the boundary
		\begin{align}
			\int_{\partial_\mu\mc B}{(\mu-\mu_{\rm e})\tilde{\eta}} = 0\,, \label{eq:weakbound}
		\end{align}	
\end{itemize}
added by the constitutive equations \eqref{eq:partialQ1}-\eqref{eq:partialQ2},\eqref{eq:stress}-\eqref{eq:chempot} and by Fick's law
\begin{align}
\mb{h}=-\mb{M}\nabla\mu\,, \quad \mb{M} = \frac{c D}{R_{\rm g}T}\mb{I}\,, \label{eq:fick}
\end{align}
where $\mb{M}$ is a positive semidefinite \textit{solvent mobility} tensor,  with $D$ the \textit{solvent diffusivity}. Here, we have assumed an isotropic mobility (with respect to the reference state)\cite{Lucantonio2013} and taken a linear dependence of $\mb{M}$ on $c$ to model changes in solvent permeability associated with changes in porosity induced by swelling. 

The relation between the weak form equations \eqref{eq:forcesweak}-\eqref{eq:weakbound} and their strong form counterparts has been described elsewhere\cite{Lucantonio2013} in the context of isotropic (non-mesogenic) polymer gels, except for eq.~\eqref{eq:weaknem} whose strong form may be readily found. We solve the system for $(\mb{u},c,p,Q,\eta)$ such that $\tilde{\eta} = 0$ on $\partial \mc{B}\setminus\partial_{\mu}\mc{B}$ using quadratic shape functions for $\mb{u}$, $c$ and $Q$, and linear shape functions for $p$ and $\eta$. 
For the numerical experiments reported in Section~\ref{sec:numerics}, the system \eqref{eq:forcesweak}-\eqref{eq:fick}, the constitutive equations and the corresponding boundary conditions, were implemented into the finite element software COMSOL Multiphysics v5.3. The implicit, variable-order (from 1 to 5), adaptive step-size BDF solver was used for the time-stepping. A quasi-Newton algorithm was employed to solve iteratively the non-linear algebraic system resulting from the finite element discretization at each time step. The direct solver MUMPS was chosen for the solution of the linearized system at each iteration. For the free swelling cube, the mesh consisted of about 256 hexahedral elements.

The governing equations of the coupled model may also be found upon imposing the stationarity of the following rate-type functional
\begin{align}
\mc{L}[\dot{\mb{F}},\dot{c},\dot{Q}, \mb{h},\lambda,\dot{p}] = \frac{\mbox{d}}{\mbox{d}t}\mc{E}[\mb{F},c,Q, p]+\mc{F}[\mb{h},\lambda,\dot{c}]+\mc{D}[\mb{h}]
\end{align}
where $\mc{E}$ is represented as in \eqref{eq:totenergy} and the term
\begin{align}
\mc{F}[\mb{h},\lambda,\dot{c}]=\int_{\partial_\mu\mc{B}}{\mu_{\rm e}\mb{h}\cdot\mb{m}}-\int_{\mc B}{\lambda(\dot{c}+\divg \mb{h})}
\end{align}
accounts for the external chemical power and enforces the balance of solvent mass through the Lagrange multiplier $\lambda$.
Here, $\mc{D}$ is the \textit{dissipation potential due to solvent transport}, which we represent as
\begin{align}
\mc{D}[\mb{h}] = \frac{1}{2}\int_{\mc B}{\mb{M}\mb{h}\cdot\mb{h}}\,.
\end{align}
The vanishing of the first variation of $\mc{L}$ with respect to $\dot{c}$ immediately identifies $\lambda$ with the chemical potential $\mu$ as given by eq.~\eqref{eq:chempot}. The constitutive equation \eqref{eq:fick} corresponds to arbitrary variations $\tilde{\mb{h}}$ in $\mc{B}$. This variational setting could provide a basis for numerical schemes based on an incremental minimization approach\cite{Boger2017}.

\section{Numerical experiments}
\label{sec:numerics}

\begin{figure*}[!b]
	\centering
	\includegraphics[scale=1.34]{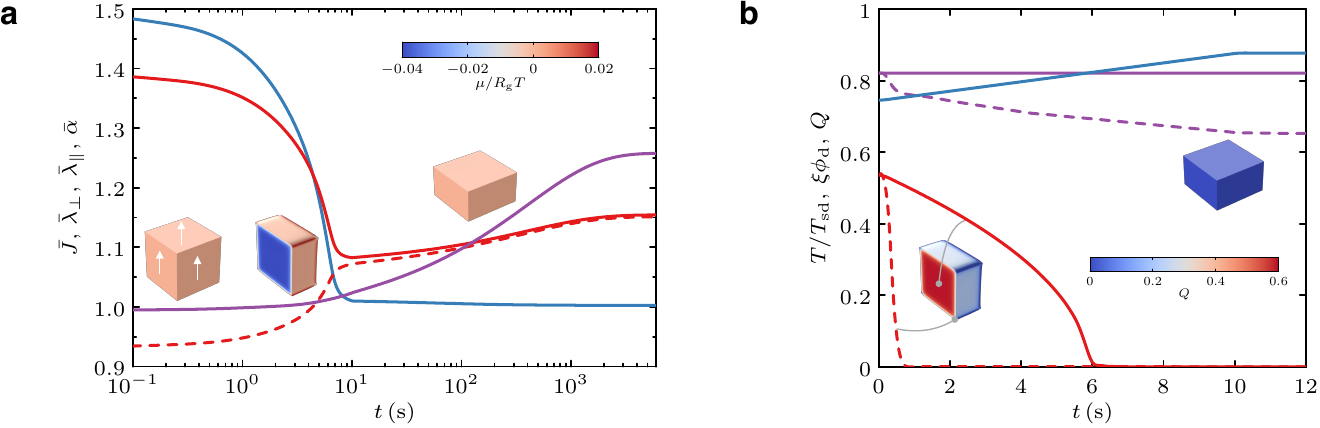}
	\caption{Concurrent free swelling and nematic-isotropic transition in a LCG. (a) Average principal stretches $\bar{\lambda}_{\parallel}$ (red solid line) and $\bar{\lambda}_{\perp}$ (red dashed line), anisotropy $\bar{\alpha} = \bar{\lambda}_{\parallel}/\bar{\lambda}_{\perp}$ (blue line), and volume ratio $\bar{J}$ (purple line) as a function of time during the N-I transition. Contour plots represent several snapshots of (a cross section of) the gel during the transient; color code is the dimensionless chemical potential $\mu/{R_{\rm g}T}$ of the solvent within the gel. Initial director orientation is marked by white arrows. (b) Time courses of the scaled temperature $T/T_{\rm sd}$ (blue), mesogens volume fraction $\xi\phi_{\rm d}$ (purple) and degree of order $Q$ (red) evaluated at the center of the cube (solid) and at a point near the  boundary (dashed). }
	\label{fig:freeswell}
\end{figure*}

To explore the coupling between swelling and nematic order, we perform a simple numerical experiment inspired by its physical counterpart reported previously\cite{Urayama2006a}. The values of the model parameters used in the numerical simulation are listed in Table~\ref{tab1} and are taken from various sources. For the dependence of the temperature $T_{\rm s}$ on swelling in eq.~\eqref{eq:A}, we choose $f(\phi_{\rm d})=\phi_{\rm d}$, \textit{i.e.}~$f$ equals the volume fraction of mesogens with respect to the dry state. We consider the free swelling of a cube of nematic gel, which initially lies in the nematic phase, in equilibrium with the external solvent, whose chemical potential is $\mu_{\rm e} = 0\ \mbox{J/mol}$. Then, the temperature of the environment is increased linearly in time from $T_{\rm o}$ at $t=0$ to $T_{\rm f} > T_{\rm sd}\xi/J_{\rm o}$ at $t=\tau=10\,\mbox{s}$, to activate the nematic-isotropic transition. The temperature-induced attenuation of the interactions among the mesogens, as quantified by the parameter $A$ in \eqref{eq:A}, leads to a reduction in the degree of order and thus in the nematic pressure, which reflects into a decrease in the chemical potential of the solvent within the gel (see the contour plots for $t>0$ in Fig.~\ref{fig:freeswell}a). In addition, the increased chemical affinity between polymer and solvent ($\chi$ decreases) at higher temperatures promotes swelling. Therefore, the chemical non-equilibrium resulting from the gap in solvent chemical potential between the gel and the environment drives solvent absorption and swelling.

\begin{table}[t]
	\centering
	\small
	\caption{\ Parameter values used in the numerical simulations.}
	\label{tab1}
	\begin{tabular*}{0.6\textwidth}{@{\extracolsep{\fill}}lll}
		\hline
		Parameter & Symbol & Value \\
		\hline
		Shear modulus of the LCE at $T=T_{\rm o}$& $G$ & $200\,\mbox{kPa}$ \\
		Solvent diffusivity & $D$ & $1\times 10^{-11}\,\mbox{m}^2/\mbox{s}$ \\
		Solvent molar volume & $\Omega$ & $1\times 10^{-5}\,\mbox{m}^3/\mbox{mol}$ \\
		Dimensionless mixing enthalpy & $\chi$ & $315\,\mbox{K}/T$ \\
		Mesogens-backbone order coupling & $a$ & 0.5\\
		Volume fraction of mesogens & $\xi$ & 0.85 \\
		Characteristic temperature of the LCE & $T_{\rm sd}$ & 447\,\mbox{K} \\
		Initial degree of order & $Q_{\rm o}$ & 0.54 \\
		Initial swelling ratio & $J_{\rm o}$ & 1.22 \\
		Initial temperature & $T_{\rm o}$ & $283\,\mbox{K}$ \\
		Final temperature & $T_{\rm f}$ & $330\,\mbox{K}$ \\
		Landau-de Gennes parameter & $A_{\rm o}$ & $1\times10^6\,\mbox{J}/(\mbox{m}^3\cdot\mbox{K})$\\
		Landau-de Gennes parameter & $B$ & $0\,\mbox{J}/\mbox{m}^3$\\
		Landau-de Gennes parameter & $C$ & $1\times10^8\,\mbox{J}/\mbox{m}^3$\\
		\hline
	\end{tabular*}
\end{table}

During the transient approach to equilibrium, both volume and shape changes occur. Since for gels thermal diffusivity $(\sim 10^{-7}\,\mbox{m}^2/\mbox{s})$ is usually several orders of magnitude greater than solvent diffusivity $(\sim 10^{-9}-10^{-11}\,\mbox{m}^2/\mbox{s})$, thermal equilibrium is reached much faster than chemical one. As a consequence, shape and volume changes, which are tied to the kinetics of phase transition and solvent absorption, respectively, occur on separate time scales, consistent with experimental observations\cite{Urayama2006a}. This separation in time-scales is quantified and clearly demonstrated in Fig.~\ref{fig:freeswell}a, where we compare the volume-averaged degree of anisotropy $\bar{\alpha} = \bar{\lambda}_{\parallel}/\bar{\lambda}_\perp$, with $\lambda_{\parallel} = \mb{F}_{\rm d}\cdot\mb{N}_{\rm o}$ and $\lambda_{\perp} = \mb{F}_{\rm d}\cdot(\mb{I}-\mb{N}_{\rm o})$ the principal stretches from the dry state, and the volume-averaged swelling degree $\bar{J}$ (a superposed bar denotes the volume average). We observe that, while the degree of anisotropy decreases and reaches $\bar{\alpha} \approx 1$ at $t \approx \tau$, volume changes become significant ($\bar{J}>1$) only for $t>\tau$.

As swelling proceeds starting from the outer layers of the sample, there is a localized dilution of the mesogens' concentration, which facilitates the disruption of the nematic order. Thus, heterogeneous swelling induces gradients in the degree of order, as reported in the plots of $Q$ in Fig.~\ref{fig:freeswell}b. Further, swelling causes a heterogeneous shift in the characteristic temperature $T_{\rm s}$, so that the transition occurs earlier in the regions that swell first. In particular, in our simulations the transition occurs approximately when the temperature equals the local value of $T_{\rm s}$ given by eq.~\eqref{eq:A}$_{2}$, and is thus marked by the intersection of the curves $T/T_{\rm sd}$ and $\xi \phi_{\rm d}$.

\paragraph*{An application to micromotility}

Having studied the coupled processes occurring during the nematic-isotropic transition, we now assess the possibility of exploiting transient shape and volume changes to actuate a prototype of microswimmer. At small length scales, viscosity-dominated swimming requires non-reciprocal shape changes to produce net motion\cite{Purcell1977}. This is a consequence of the time-reversibility of the Stokes' equations, which govern the flow in the low-Reynolds-number regime. Graphically, non-reciprocal shape changes correspond to loops in the space of shape parameters\cite{Alouges2008,Alouges2009}.

Among the various minimal microswimmers that have been proposed to fulfill these theoretical requirements for locomotion, here we study the ``pushmepullyou'' (PMPY). In its original flavor\cite{Avron2005}, this swimmer consists of two spheres connected by a rod. The spheres can change their volumes in an alternating fashion, \textit{i.e.} such that the total volume is conserved, whereas the rod can vary its length. In the variant we propose, the rod and one of the spheres are made of a LCG initially ($T=T_{\rm i}$) in its nematic phase, with $\mb{N}_{\rm o}$ oriented perpendicularly with respect to the longitudinal axis of the rod, while the other sphere is rigid. As the rod stretches, its length becomes $\lambda_{\perp}(t)L$, where $L$ is the initial length and $\lambda_{\perp}$ is the stretch in the direction perpendicular to the nematic orientation, \textit{i.e.} the longitudinal axis of the rod. We indicate with $R_{i}(t)$ the radius of the $i$-th sphere, $i=1,2$, with $i=1$ corresponding to the LCG sphere, and we set, for simplicity, $R_1(0)=R_2(0)=R_{\rm o}$. We denote by $J(t)$ the swelling ratio of the LCG sphere, so that the ratio between the current radius $R_1(t)$ at time $t$ and the reference one $R_1(0)$ is $J^{1/3}(t)$.

\begin{figure}[t]
	\centering
	\includegraphics[scale=1.22]{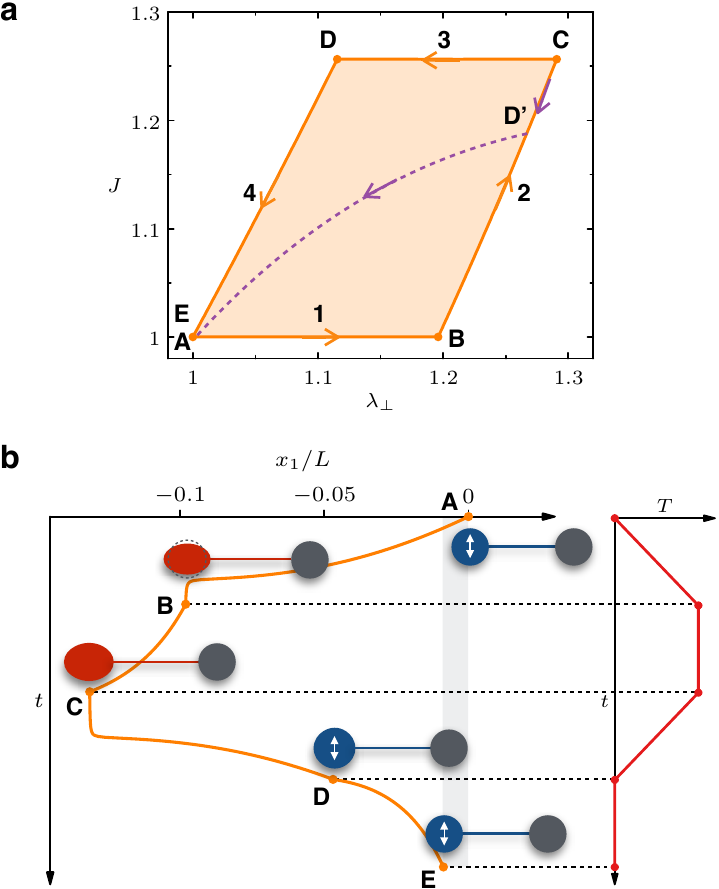}
	\caption{Pushmepullyou nematic gel swimmer. A loop in the swelling ratio-longitudinal stretch ($J,\lambda_{\perp}$) plane (a) allows for net motion of the microswimmer in Stokes flow. (b) Scaled position $x_1/L$ of the left sphere as a function of time and sketch of the 4 phases characterizing a swimming stroke. The left sphere and the rod joining the spheres are made of a LCG, while the right sphere (grey) is rigid. Director orientation is shown as a white arrow. Red and blue colors identify the isotropic and nematic phases of the LCG, respectively. Shaded area corresponds to the net motion at the end of one stroke. On the right, the temperature profile is reported.}
	\label{fig:swimmerloop}
\end{figure}

By exploiting the separation in the time scales of volume and shape changes, we can design a loop in the space of controls $(J,\lambda_{\perp})$, as depicted in Fig.~\ref{fig:swimmerloop}a, which is characterized by the following phases:
\begin{enumerate}
	\item shape change (constant volume) induced by the N$\rightarrow$I phase transition upon increasing the temperature from $T=T_{\rm i}$ to $T=T_{\rm max}>T_{\rm ni}$;
	\item isotropic swelling at constant temperature $T=T_{\rm max}$;
	\item shape change (constant volume) induced by the I$\rightarrow$N phase transition upon decreasing the temperature back to $T=T_{\rm i}$;
	\item anisotropic shrinking at constant temperature $T=T_{\rm i}$.
\end{enumerate}
Here, for simplicity, we study the actuation cycle as a sequence of homogeneous equilibrium states. Hence, time plays the role of a load parameter that we choose such that each step has unit duration. During the cycle the external chemical potential is held fixed to its initial value $\mu_{\rm o}$, which determines the reference state (A). The configurations of the system corresponding to the loop in Fig.~\ref{fig:swimmerloop}a are sketched in Fig.~\ref{fig:swimmerloop}b, along with the temperature profile. The transformation A-B can be characterized by solving the system eq.~\eqref{eq:stressfree1}-\eqref{eq:stressfree2},\eqref{eq:freeswellQ} with the isochoric constraint $J=\lambda_{\parallel}\lambda_{\perp}^2=1$, while $T$ is ramped linearly in time from $T_{\rm i}$ to $T_{\rm max}>T_{\rm ni}$.  Chemical non-equilibrium drives swelling in the transformation B-C. Indeed, the reduction in the modulus of the nematic pressure associated with the N$\rightarrow$I transition occurring in A-B causes the chemical potential to drop below $\mu_{\rm o}$, as it results from eq.~\eqref{eq:chempot},\eqref{eq:Pifree}-\eqref{eq:pnfree} with $Q=0$ and the pressure $p$ given by eqs.~\eqref{eq:stressfree1}-\eqref{eq:stressfree2}. 
The intermediate states that belong to the transformation B-C may be determined by solving the system of equations \eqref{eq:stressfree1}-\eqref{eq:freeswellQ} including the chemical equilibrium condition with $\mu_{\rm e}$ increasing from the non-equilibrium value of state B to $\mu_{\rm o}$, which is attained at C. 

Phases 3 and 4 (C-D and D-E), \textit{i.e.}~the recovery stroke, deserve particular attention, since their sequence is crucial for the non-reciprocal character of the actuation cycle. In the scenario corresponding to Fig.~\ref{fig:swimmerloop}a, volume changes (D-E) follow shape changes (C-D), thus they mirror the transformations A-B and B-C (the power stroke) and are ruled by analogous governing equations. This case holds when the temperature decrease in C-D is sufficient to trigger the I$\rightarrow$N transition. However, if the dilution of nematic interactions due to swelling after phase 2 is strong enough to lower the transition temperature below $T_{\rm i}$, the gel remains in the isotropic phase as $T$ decreases in phase 3. In this case, isotropic shrinking starts \textit{before} the I$\rightarrow$N transition, as triggered by the increase of the chemical potential above $\mu_{\rm o}$. Hence, during the transformation C-D the system moves back along B-C, until the I$\rightarrow$N transition occurs (D') when $T$ meets the value of $T_{\rm s}$, which continuously increases as shrinking proceeds. From D' back to the original state (E), concurrent volume and shape changes take place. In the plane $(J,\lambda_{\perp})$ this case corresponds to a ``triangular'' loop, whose area and thus net motion per cycle would be smaller than those for the ``parallelogram'' cycle described above, where shape and volume changes are time-separated in both the power and the recovery strokes. Intuitively, the type of recovery stroke between the two alternatives just illustrated is related to the dependence of $T_{\rm s}$ on swelling, \textit{i.e.}~to the function $f(\phi_{\rm d})$. In particular, we have found that the power-law $f(\phi_{\rm d})=\phi_{\rm d}^{1/4}$ with a weaker dependence on $\phi_{\rm d}$ with respect to a linear relation allows to perform a ``parallelogram''-type stroke, given the values of the other model parameters.

Once the time-histories $J(t)$ and $\lambda_{\perp}(t)$ of the controls are assigned, we can compute the flow field around the swimmer and, then, the positions of the spheres in time. For this purpose, we assume that the radii of the spheres are much smaller then the distance between them ($R_i/L\ll 1$), so that the solution can be constructed from the one for a single sphere using the linearity of the Stokes equations for low Reynolds number hydrodynamics. Following this approach, to the leading order in $R_i/L$, the velocity of the $i$-th sphere whose center is located at $x_i$ can be written as\cite{Avron2005}
\begin{align}
	\dot{x}_{i} = -\frac{F_i}{6\pi\epsilon R_i}+ \frac{ \dot{V}}{4\pi(\lambda_{\perp}L)^2}\,, \quad \dot{\lambda}_{\perp} = \frac{\dot{x}_2 - \dot{x}_1}{L}\,. \label{eq:position1}
\end{align}
The first term in eq.~\eqref{eq:position1}$_1$ is the viscous drag contribution according to Stokes' law, with $\epsilon$ the solvent viscosity and $F_i(t)$ the force acting on the $i$-th sphere. The second term accounts for the change in the volume $V(t)$ of the left sphere. In the ensuing calculations, we neglect the drag acting on the rod and we estimate $R_1$ as the radius of a sphere having the same volume as that of the ellipsoid, when the LCG undergoes the N$\rightarrow$I transition, so that $V=4\pi R_1^3/3$. Upon imposing that the force resultant $F_1 + F_2$ acting on the swimmer vanishes, eq.~\eqref{eq:position1} may be recast in dimensionless form as
\begin{align}
\frac{\dot{x}_1}{L} = -\frac{1}{1+J^{1/3}}\dot{\lambda}_{\perp}+\left(\frac{R_{\rm o}}{L}\right)^3 \frac{\dot{J}}{3\lambda_{\perp}^2}\,, \quad \frac{\dot{x}_2}{L} = \frac{\dot{x}_1}{L} + \dot{\lambda}_{\perp}\,. \label{eq:position}
\end{align}
Notice that the second contribution in eq.~\eqref{eq:position}$_1$ is much smaller than the first.
Integration of eqs.~\eqref{eq:position} using the functions $J(t)$ and $\lambda_{\perp}(t)$ that we  computed to define the actuation cycle in Fig.~\ref{fig:swimmerloop}a provides the position $x_1(t)$ reported in Fig.~\ref{fig:swimmerloop}b. With the values of the parameters we have chosen and $R_{\rm o}/L = 0.1$, the swimmers moves in each cycle by approximately $1\%$ of its original length. This value is rather small, but is consistent with previously reported\cite{Alouges2009} values for comparable longitudinal stretches and volume changes. 
This result should be compared with similar calculations for liquid crystal elastomer crawlers, \textit{i.e.}~LCE strips exploiting directional frictional contact with a solid surface\cite{DeSimone2015}. In fact, LCE crawlers can exploit the available maximal elongation in a cycle more effectively than LCG swimmers, reaching displacement per cycle of the order of the maximal available deformation (see equations (4.19) and (5.40) in Ref.~6), which is in turn a sizable fraction of the body length if deformations are sufficiently large.
	
The displacement per cycle could be increased by enhancing the magnitude of the swelling-induced deformations, by making the gel more compliant, for instance. However, the greater dilution associated to larger swelling would tend to make the stroke ``triangle''-like, as explained above, thus limiting the maximum achievable motion per stroke.

Finally, several strategies could be employed to reduce actuation times, another critical factor for the overall swimming efficiency that has not been taken into account in the present analysis. For instance, slow diffusive processes may be accelerated by scaling down system size. Furthermore, the temperature of the LCG may be rapidly increased via photo-thermal heating\cite{Hauser2016}, \textit{i.e.} by embedding nanoparticles in the polymer matrix to efficiently convert light into heat.

\section{Conclusions}

We have established a coupled, three-dimensional theory of swelling nematic gels accounting for network elasticity, solvent migration and liquid-crystal type interactions among the mesogens. We have also presented a variational formulation of the theory allowing for its finite element implementation. By studying the transient free swelling of a LCG undergoing a nematic-isotropic transition, we have elucidated the coupling between swelling and nematic order and the separation in time scales associated to shape and volume changes. Further, we have shown that such a separation may be exploited for  swimming  in the low-Reynolds number regime, where it translates into the non-reciprocity of shape changes that is mandatory to achieve net motion.
In general, our results on the coupled effects that characterize LCGs suggest possible applications in shape morphing and \textit{smart} sensing/actuation. 


\section*{Acknowledgments}
The authors acknowledge support from the European Research Council (AdG-340685--MicroMotility). AL also received partial support from GNFM-INdAM through the initiative ``Progetto Giovani''. AL thanks Kenji Urayama, Alfio Grillo, Luciano Teresi and Giovanni Noselli for useful discussions.


\providecommand*{\mcitethebibliography}{\thebibliography}
\csname @ifundefined\endcsname{endmcitethebibliography}
{\let\endmcitethebibliography\endthebibliography}{}

\end{document}